\documentclass[twocolumn,prl,showpacs,preprintnumbers,amsmath,amssymb]{revtex4-1}

\usepackage{graphicx}
\usepackage{verbatim}
\usepackage{color}

\begin{document}


\title{Cavity-based single atom preparation and high-fidelity hyperfine state readout}

\author{Roger Gehr}
\author{J\"urgen Volz}
\author{Guilhem Dubois}
\author{Tilo Steinmetz}
\altaffiliation{Present address:  Menlo Systems GmbH, Germany }
\author{Yves Colombe}
 \altaffiliation{Present address: NIST, Boulder, CO-80305, USA }
\author{Benjamin L. Lev}
 \altaffiliation{Present address: University of Illinois, IL-61801, USA  }
\author{Romain Long}
\author{J\'er\^ome Est\`eve}
\author{Jakob Reichel}
\email{jakob.reichel@ens.fr}

\affiliation{%
Laboratoire Kastler-Brossel, ENS, CNRS, UPMC, 24 rue Lhomond, 75005 Paris, France}%

\date{\today}
\begin{abstract}
We prepare and detect the hyperfine state of a single $^{87}$Rb atom coupled to a fiber-based high finesse cavity on an atom chip. The atom is extracted from a Bose-Einstein condensate and trapped at the maximum of the cavity field, resulting in a reproducibly strong atom-cavity coupling. We use the cavity reflection and transmission signal to infer the atomic hyperfine state with a fidelity exceeding 99.92\% in a read-out time of 100\,$\mu$s. The atom is still trapped after detection.
\end{abstract}

\pacs{42.50.Pq, 42.50.Dv, 67.85.Hj}
\maketitle


\newcommand{\ket}[1]{\left|#1\right\rangle}

A single neutral atom with two hyperfine ground states provides a long-lived two-level system ideally suited for quantum information purposes. The collisional interaction between two atoms in the vibrational ground state is a powerful mechanism for the creation of entanglement in this system \cite{Jaksch1999,Treutlein2006a}. This has been demonstrated for atoms in the Mott insulator state in an optical lattice loaded from a Bose-Einstein condensate (BEC) \cite{Mandel2003, Anderlini2007}. However, single-site addressability is challenging in these experiments \cite{Bakr2009}. Bottom-up approaches starting with laser-cooled single atoms in easily addressable macroscopic traps \cite{Schrader2004, Volz2006, Urban2009, Wilk2010} and on atom chips \cite{Teper2006} have not yet succeeded in ground state preparation. Furthermore, read-out of the qubit state is usually destructive and does not fulfill the requirements for efficient quantum error correction \cite{Knill2001}.

In this letter, we follow an intermediate route in which a single atom is extracted from a BEC trapped inside a fiber-based high-finesse cavity on an atom chip \cite{Colombe2007}. This minimizes thermal fluctuations of the extracted atom and allows us to achieve high-fidelity read-out of the hyperfine state. An atom-cavity system in the strong coupling regime enables the efficient preparation and detection of a single atom because the presence of one atom in the hyperfine state resonant with the cavity drastically changes cavity reflection and transmission \cite{Haroche2006}. Since we can trap the entire BEC in a single antinode of the intracavity dipole trap \cite{Colombe2007}, the atom position along the cavity axis is not subject to uncertainties that lead to variations in the coupling strength \cite{Boca2004,Maunz2005a,Khudaverdyan2008}. Together with the high cooperativity of our system, this allows us to reach a read-out fidelity exceeding 99.92\% in a detection time of 100\,$\mu$s without the loss of the atom \footnote{During the process of submitting this paper, we became aware of related work by Bochmann et.al., arXiv:1002.2918v1}, on a par with ion trap experiments \cite{Hume2007, Myerson2008}, the best qubit detectors so far.
 Furthermore, our cavity-assisted read-out scheme is intrinsically faster than free-space fluorescence measurements and detection times down to the sub-microsecond range are possible.

\begin{figure}
	\centering
	\includegraphics[width=\columnwidth]{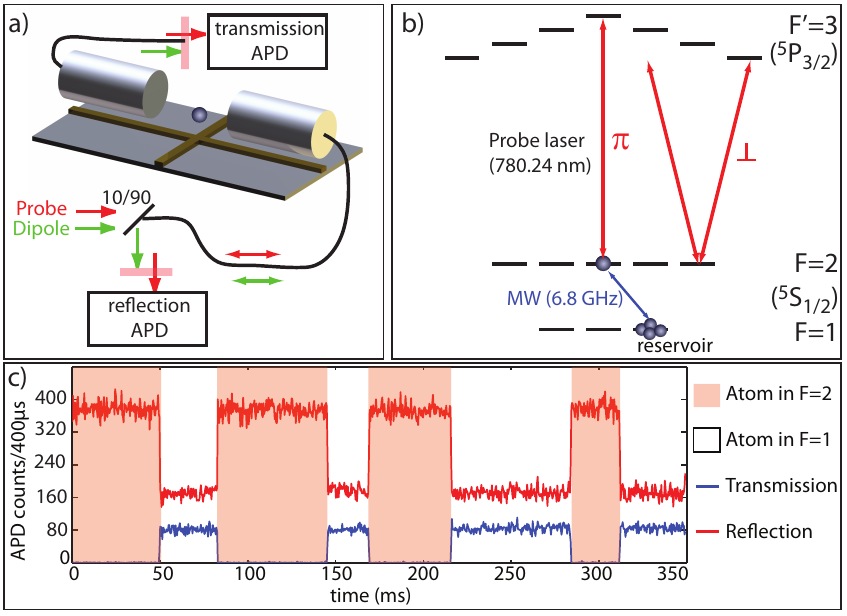}
	  \caption{\label{fig:scheme} a) Schematic of the experiment. A fiber-based high-finesse cavity is doubly resonant to a dipole trap laser and a probe laser. Two APDs record cavity reflection and transmission. b) Relevant level scheme of a single $^{87}$Rb atom trapped in the linearly $\pi$ polarized dipole trap. The cavity sustains two modes with orthogonal polarizations, $\pi$ and $\perp$. We use probe light near resonant to the $F=2\rightarrow F'=3$ transition of the D2 line. A single atom is extracted from a small reservoir using weak MW pulses. c) Typical experimental trace of cavity reflection and transmission with a single $^{87}$Rb atom coupled to the cavity. Quantum jumps between the hyperfine ground states lead to sudden, simultaneous changes in both signals.}
\end{figure}

Our experimental setup, shown in Fig.~\ref{fig:scheme}, is similar to the one described in Ref. \cite{Colombe2007}. The centerpiece of the experiment is a Fabry-Perot cavity mounted on an atom chip. The cavity with finesse 38000 and linewidth of $\kappa/2\pi$ = 53\,MHz is formed by the concave endfacets of two optical fibers with high-reflectivity coatings. The calculated maximum coupling strength between a single $^{87}$Rb atom and the cavity field is $g_{0}/2\pi$=215\,MHz for the $\ket{F=2,m_{F}=\pm 2}\rightarrow \ket{F'=3,m_{F}=\pm 3}$ transition of the $D_{2}$ line. Together with the $^{87}$Rb atomic decay rate $\gamma/2\pi$ = 3\,MHz, this positions our system in the strong coupling regime ($g_{0}\gg\kappa,\gamma$). Two avalanche photodiodes (APDs) record transmission and reflection of a probe laser. The length of the cavity is stabilized by a Pound-Drever-Hall setup using a laser at 830 nm which also serves as an intracavity standing wave dipole trap. We have determined the dipole trap-induced light shift by measuring the axial dipole trap frequency (960\,kHz). A laser beam on the repump transition is aligned transversally to the cavity axis.

The birefringence of the cavity induces a splitting of $\Delta/2\pi$=540\,MHz between its two linearly polarized eigenmodes. The dipole trap polarization is parallel to the polarization of the higher frequency cavity mode and defines the quantization axis. The pumped cavity mode is near-resonant to the atomic transition frequency, $\omega_{c}=\omega_{a}+\Delta_{ca}$, where $\omega_{a}$ is the $| F=2,m_{F}=0\rangle\rightarrow |F'=3,m_{F}=0\rangle$ transition frequency corrected for the dipole trap induced light shift of 95\,MHz, and $\Delta_{ca}$ is the cavity-atom detuning.

In a typical sequence, we prepare a small BEC of 600 atoms in a magnetic trap. We then move the trap close to one of the cavity mirrors to reduce the number of trapped atoms by surface evaporation.
After this evaporation, we are left with a small reservoir of less than 10 atoms. We load it into the central antinode of the intracavity dipole trap~\cite{Colombe2007}, ramp down the magnetic trap and apply a homogeneous magnetic field of 3.7\,G. We then use the following method to prepare a single atom. A microwave-induced adiabatic rapid passage transfers all atoms to the $\ket{F=1, m_{F}=1}$ state, where they act as a dispersive medium for the cavity, shifting its resonance by -6.1\,MHz per atom. A cavity transmission measurement allows us to approximately deduce the number of atoms in the reservoir. We post-select only runs where transmission is compatible with less than 6 atoms in the reservoir, since a larger reservoir later increases the probability to extract two rather than one atom.

\begin{figure}
	\centering
		\includegraphics[width=\columnwidth]{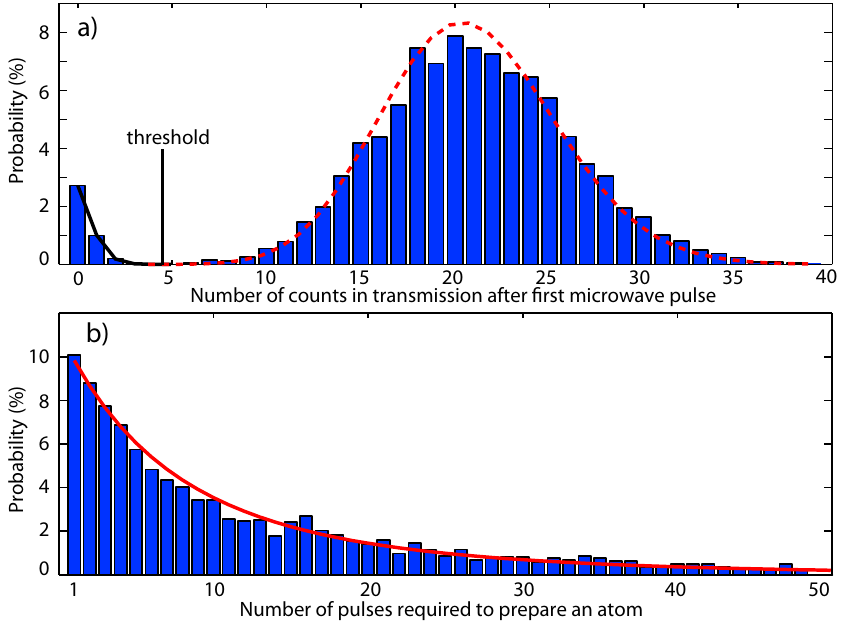}
	  \caption{\label{fig:preparation} a) Histogram of counts in transmission during 20\,$\mu$s after the first microwave pulse. The full black and dashed red lines are poissonian fits with expectation values of 0.3 and 22 respectively. The clear dip between the two peaks allows us use a thresholding technique to signal the preparation of a single atom. b) Distribution of the number of required MW pulses to prepare an atom. The line is a fit assuming a reservoir with a poissonian distribution.}
\end{figure}

We now apply a 1.9\,$\mu$s microwave (MW) pulse resonant to the $\ket{F=1,m_{F}=1}\rightarrow \ket{F=2,m_{F}=0}$ transition with a transfer probability of $4.2\%$ per atom. In order to detect a successful transfer, we measure the cavity transmission during 20\,$\mu$s with a probe laser resonant to cavity and atom. Figure~\ref{fig:preparation}a) shows the histogram of detected counts on the transmission APD following the MW pulse. The measured probability distribution is well approximated by the sum of two poissonian distributions. The low transmission peak corresponds to the presence of at least one $F=2$ atom, whereas the high transmission indicates that no atom was transferred. The distribution drops close to zero between the two peaks, justifying the choice of a threshold at 5 counts to determine whether an atom was transferred. The probability of a false positive event, i.e. a drop of cavity transmission below threshold although no atom is in $F=2$, is negligible $(\approx10^{-5})$.

We repeatedly apply this preparation cycle until a transmission level below threshold signals a successful transfer. Figure~\ref{fig:preparation}b) shows the probability distribution of the number of pulses required to prepare an atom. Sequences in which no transfer occurs after 50 trials are discarded. A poissonian reservoir atom number distribution with mean atom number $\overline{n}=1.5$ fits the data well. This measurement allows us to quantify the quality of the single atom preparation. We calculate the probability that a successful preparation leads to more than one atom to be 2.6\%.

Since we want to detect the hyperfine state of a single atom, we now remove the remaining atoms in $F=1$. We do this by lowering the dipole trap and turning on a magnetic field gradient of 2.7\,kG$/$cm during 30\,ms, creating a strong force on the reservoir in state $\ket{F=1,m_{F}=1}$ that exceeds the restoring force of the dipole trap. The single atom in $F=2$ has a strong probability to be in the magnetic field-insensitive state $m_{F}=0$ and remains trapped, while in approximately 99$\%$ of the runs all reservoir atoms are removed. After this procedure, a probe laser pulse verifies that the atom is still trapped. 

In order to show that all single atoms prepared in this way strongly couple to the cavity with similar strength, we measure the normal-mode spectrum of the atom-cavity system~\cite{Maunz2005a,Boca2004}. We probe cavity transmission at a given laser-cavity detuning $\Delta_{lc}$ during 8\,$\mu$s. To ensure that the atom is still trapped, we then apply a short repump pulse and check that the on-resonance cavity transmission is below the preparation threshold. This measurement-control cycle is repeated until the atom is lost. The resulting cavity transmission for a given detuning is obtained by averaging over approximately 20 atoms.

\begin{figure}
	\centering
		\includegraphics[width=\columnwidth]{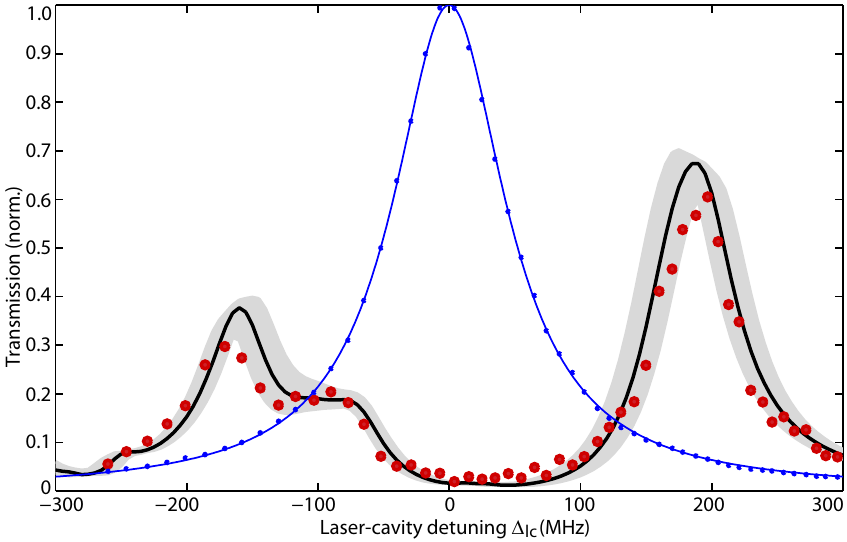}
	  \caption{\label{fig:rabi} Normal mode spectrum of a single atom coupled to the cavity. Large red points are the measured transmission. The thick line is the steady-state solution of the master equation for $g_{0}/2\pi=240$\,MHz and $\Delta_{ac}=0$. The gray area bounds the solutions when varying $g_{0}/2\pi$ by $\pm 10$\,MHz and $\Delta_{ac}/2\pi$ by $\pm 10$\,MHz. The small blue dots are the measured empty cavity transmission together with a Lorentzian fit.}
\end{figure}

The resulting normal-mode spectrum is shown in Fig.~\ref{fig:rabi} together with the steady state prediction of the atom-cavity master equation. In order to account for all features of the spectrum, the model contains the full Zeeman structure of the $F=$2 and $F'=1,2,3$ manifolds as well as the magnetic field, dipole trap light shift, and coupling to both cavity modes. The probe light pumps the $\pi$-polarized higher frequency cavity mode. The coupling strength $g_{0}$ is the only free parameter in the model.

The value of $g_{0}$ extracted from the fit is $240\pm10$\,MHz, 12$\%$ higher than expected from the calculated cavity mode geometry and whithin the error given by the uncertainty in the curvatures of our mirrors. The high value of the observed coupling and the absence of peak broadening are strong indications that a large fraction of the prepared atoms is well localized close to the maximum of the cavity field. This is in agreement with the observed binary transmission level during preparation (see Fig.~\ref{fig:preparation}): All atoms extracted from the ultracold reservoir couple maximally to the cavity.

 The high degree of control on the external degree of freedom of the prepared atom is an ideal starting point to establish the performance of the cavity as a qubit detector. The fidelity of the read-out is defined as $\mathcal{F}=1-\epsilon$, where $\epsilon=(\epsilon_{F1}+\epsilon_{F2})/2$, and $\epsilon_{F2}$ ($\epsilon_{F1}$) is the probability of detecting $F=1$ ($F=2$) if the atom is in $F=2$ ($F=1$) immediately before the measurement. The atomic state is inferred from the cavity reflection and transmission signals with the laser resonant to the cavity. A combination of low (high) transmission and high (low) reflection signals an atom in $F=2$ ($F=1$). A typical trace is shown in Fig.~\ref{fig:scheme}c). In this trace, the atom is probed for a time much longer than needed for detection, and the atom performs quantum jumps between the two hyperfine ground states under the action of probe light. We implement two methods to infer the atomic state from the registered counts in reflection and transmission.
The first method integrates both count rates for a given detection time and uses thresholding in the two-dimensional space $\mathcal{C}$ of all possible counts in reflection and transmission. The second method additionally makes use of the temporal evolution of the signal for a maximum likelihood estimation of the atomic state.

The detection efficiency of both schemes is determined by the lifetimes of the hyperfine states and the different count rates for the two atomic states. The hyperfine lifetime is limited by optical pumping caused by the probe light. We optimize all parameters, count rates and lifetimes, by pumping the $\perp$-polarized cavity mode, by reducing the magnetic field to 1\,G and by setting $\Delta_{ca}=-20$\,MHz. These settings lead to efficient optical pumping into the extremal Zeeman states $|F=2,m_{F}=\pm2\rangle$, see Fig.~\ref{fig:scheme}b).

Traces like the one shown in Fig.~\ref{fig:scheme}c) allow the direct measurement of both count rates and lifetimes. Individual lifetimes are distributed exponentially. A fit to the distribution gives an average lifetime of $\tau_{F2}$=52 ms ($\tau_{F1}$=26 ms) for $F=2$ ($F=1$) atoms at the cavity pump rate of $1.9\cdot10^{6}$ photons/s. The count rates in transmission (reflection) of the cavity with an atom in $F=2$ are $1.4\cdot10^{3}$/s ($8.9\cdot10^{5}$/s), with an atom in $F=1$ they become $1.9\cdot10^{5}$/s ($4.4\cdot10^{5}$/s). Optical losses and 60$\%$ APD detection efficiency account for the difference in pump power and detected flux.

\begin{figure}[t]
  \centering
		\includegraphics[width=\columnwidth]{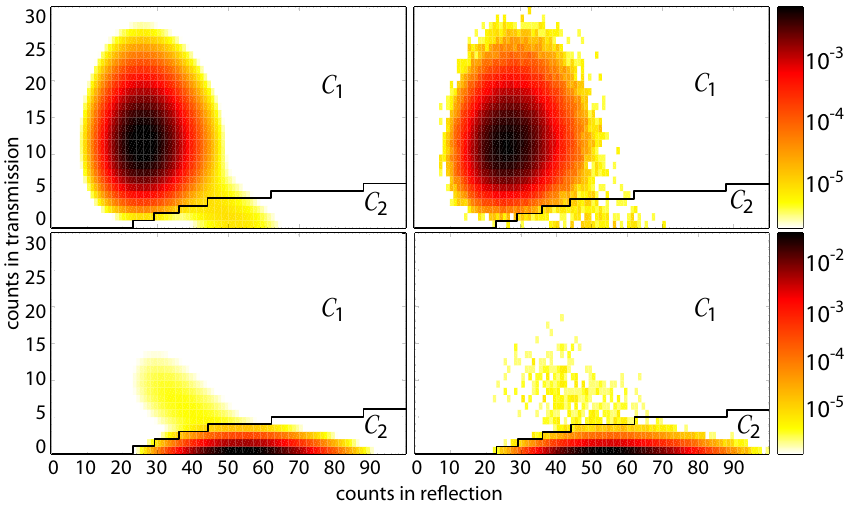}
		\caption{\label{fig:detection} Two dimensional distributions of counts registered during 60\,$\mu$s in reflection and transmission with an atom prepared in $F=1$ (upper graphs) and $F=2$ (lower graphs). The left graphs are calculated distributions. The right graphs show the associated measurements. The colorscale logarithmically encodes $p(c_{R},c_{T})$. The solid black lines represent the threshold used to differentiate $F=2$ from $F=1$ atoms.}
\end{figure}

For the thresholding method, we calculate $p_{F2}(c_{R},c_{T})$ ($p_{F1}(c_{R},c_{T})$) which is the probability distribution to observe $(c_{R},c_{T})$ counts in reflection and transmission for a given detection time when the atom initially is in state $F=2$ ($F=1$). The model assumes exponential decays of the hyperfine states and poissonian distributions of detected counts. Outcomes in the subspace $\mathcal{C}_{2}$ defined by the threshold $p_{F2}(c_{R},c_{T}) > p_{F1}(c_{R},c_{T})$ signal an atom in $F=2$ and vice versa. The detection errors are given by $\epsilon_{F1}$=$\sum_{\mathcal{C}_{2}} p_{F1}$ and $\epsilon_{F2}$=$\sum_{\mathcal{C}_{1}} p_{F2}$. For short integration times, these errors are dominated by photon shot noise. Increasing the integration time decreases shot noise, but the probability of a quantum jump during detection increases. An optimum detection time exists, for which the sum of both errors is minimized. Figure~\ref{fig:detection} shows the two calculated counts distributions for the optimum detection time of 60\,$\mu$s.

The simple count thresholding discards the useful information encoded in the temporal evolution of the signal. A maximum likelihood method circumvents this problem \cite{Myerson2008}. Each detection pulse is divided into $N$ time bins. The single outcome $(c_{R},c_{T})$ is replaced by the set $M$=\{$c_R^i,c_T^i$\}, where $i=1...N$ refers to the bin number. For each set $M$ we calculate $q_{F2}(M)$ ($q_{F1}(M)$), the probability of obtaining the data set $M$ if the initial state is $F=2$ ($F=1$). If $q_{F2}(M)\geq q_{F1}(M)$, we conclude that the atom was in $F=2$ before the detection and in $F=1$ otherwise. Both probabilities are calculated recursively by considering more and more bins. The maximum number of bins $N$ is chosen such that a further increase in detection time does not significantly change the outcome. For our parameters, this results in a measurement time of 100\,$\mu$s. This method generates a better state inference for atoms that change their state during the detection time.

In order to experimentally determine the errors of both read-out methods, we have to prepare a single atom in a well-defined hyperfine state before applying either detection method. For this, we load a single atom into the dipole trap and continuously monitor cavity transmission until the atom is lost. The observation of transmission below a lower (above a higher) threshold signals a succesful preparation in $F=2$ ($F=1$), and starts a detection. Using two separate thresholds reduces the uncertainty of the prepared state at the price of discarding results with intermediate transmission \footnote{With the chosen thresholding parameters, only about 10\% of the data is discarded. A hypothetical bias introduced by this method can therefore only have a small influence on the measured detection fidelity.}. The state preparation fidelity is limited by the finite probability of a quantum jump occuring during the preparation time, leading to errors of 2.4$\cdot 10^{-4}$ (1.2$\cdot 10^{-4}$) in the preparation of $F=2$ ($F=1$) atoms. We emphasize that we observe no time-dependence of the cavity transmission with an $F=2$ atom, except for the few microseconds preceding the loss of the atom from the trap. This allows us to use the whole time during which the atom is trapped for repeated preparation and detection. On average 1000 preparation-detection cycles are performed on each atom.

Table~\ref{table1} shows the measured and calculated detection errors. As expected, the maximum likelihood method gives the best read-out fidelity of 99.92\%. The measured thresholding method fidelity is 99.9\%, very close to the prediction of the model. The difference is mostly accounted for by state preparation errors, which decrease the measured read-out fidelity. Accidental preparation of more than one atom only affects the measured detection error on the level of $10^{-5}$. 

Fig.~\ref{fig:detection} shows the comparison between the calculated and measured probability distributions, from which we extract the errors for the thresholding method. The tails of the distributions that make up the dominant contribution to the read-out error are clearly visible on both the measured data and the calculation. They are caused by a quantum jump occuring during the detection time.

The single atom preparation and high-fidelity read-out presented in this paper constitute two major steps towards single atom quantum engineering on atom chips. Together with recently realized state-dependent microwave potentials \cite{Boehi2009}, all individual elements for chip-based two-qubit gates have now been demonstrated.
Other applications may also benefit from the features of our read-out scheme and in particular its high bandwidth. Increasing probe power, we have measured 99.4$\%$ fidelity in 2\,$\mu$s, limited by APD dead time which becomes important for count rates exceeding 5\,MHz. At even higher probe powers, the dynamics of the atom-cavity system becomes non-linear; the fundamental limit is only reached for a detection time on the order of $1/\kappa$, which for our system is 3\,ns.
Sub-microsecond detection times can be useful for many experiments, e.g. the realization of loophole-free Bell tests \cite{Rosenfeld2009}. Additionally, the residual light intensity inside the cavity is weak when the atom is in the state resonant to the probe light. Compared to fluorescence measurements, the amount of spontaneous emission is thus greatly reduced. This opens up the possibility of an ideal projective measurement of the qubit state. 

\begin{table}[t]
	\centering
		\begin{tabular}{|p{0.5cm}|p{1.5cm}|p{1.5cm}|p{1.5cm}|p{1.5cm}|p{1.1cm}|}	  
		\hline
		                            & \centering TM               & \centering TM                 &\centering MLM       &\centering MLM   &\centering prepa- \tabularnewline           
		 
		 & \centering calculated                           & \centering measured                 &\centering calculated       &\centering measured  &\centering  ration\tabularnewline
		                          
		 \hline
		 $\centering \epsilon_{F1}$ &  \centering $7.0$ &\centering $9.6\pm0.6$ &\centering $4.8$ &\centering $8.7\pm0.6$ &\centering $1.2$\tabularnewline
		 \hline
		 $\centering \epsilon_{F2}$ &  \centering 9.1 &\centering $10.9\pm0.5$ &\centering $4.9$ &\centering $7.2\pm0.4$  &\centering $2.4$\tabularnewline
		 \hline \hline
		 $\centering \epsilon$    &  \centering   8.0 &\centering $10.3\pm0.4$ &\centering $4.9$ &\centering $7.9\pm0.3$  & \tabularnewline
		 \hline 
		\end{tabular}
		\caption{\label{table1} Calculated and measured hyperfine state read-out errors for the thresholding method (TM) and the maximum likelihood method (MLM). All numbers have to be multiplied by $10^{-4}$. Uncertainties are statistical. The last colunm gives preparation errors.}
\end{table}

We gratefully acknowledge financial support for this work from a EURYI award and the SCALA Integrated Project of the EU.


\begin{thebibliography}{10}

\bibitem{Jaksch1999}
D. Jaksch {\it et~al.}, Phys. Rev. Lett. {\bf 82},  1975  (1999).

\bibitem{Treutlein2006a}
P. Treutlein {\it et~al.}, Phys. Rev. A {\bf 74},  022312  (2006).

\bibitem{Mandel2003}
O. Mandel {\it et~al.}, Nature {\bf 425},  937  (2003).

\bibitem{Anderlini2007}
M. Anderlini {\it et~al.}, Nature {\bf 448},  452  (2007).

\bibitem{Bakr2009}
W.~S. Bakr {\it et~al.}, Nature {\bf 462},  74  (2009).

\bibitem{Schrader2004}
D. Schrader {\it et~al.}, Phys. Rev. Lett. {\bf 93},  150501  (2004).

\bibitem{Volz2006}
J. Volz {\it et~al.}, Phys. Rev. Lett. {\bf 96},  030404  (2006).

\bibitem{Urban2009}
E. Urban {\it et~al.}, Nature Phys. {\bf 5},  110  (2009).

\bibitem{Wilk2010}
T. Wilk {\it et~al.}, Phys. Rev. Lett. {\bf 104},  010502  (2010).

\bibitem{Teper2006}
I. Teper, Y.-J. Lin, and V. Vuletic, Phys. Rev. Lett. {\bf 97},  023002
  (2006).

\bibitem{Knill2001}
E. Knill, R. Laflamme, and G.~J. Milburn, Nature {\bf 409},  46  (2001).

\bibitem{Colombe2007}
Y. Colombe {\it et~al.}, Nature {\bf 450},  272  (2007).

\bibitem{Haroche2006}
S. Haroche, and J.-M. Raimond, Exploring the Quantum, Oxford Univ. Press (2006).


\bibitem{Boca2004}
A. Boca {\it et~al.}, Phys. Rev. Lett. {\bf 93},  233603  (2004).

\bibitem{Maunz2005a}
P. Maunz {\it et~al.}, Phys. Rev. Lett. {\bf 94},  033002  (2005).

\bibitem{Khudaverdyan2008}
M. Khudaverdyan {\it et~al.}, New Journal of Physics {\bf 10},  073023
  (2008).

\bibitem{Note1}
During the process of submitting this paper, we became aware of related work by
  Bochmann et.al., arXiv:1002.2918v1.
  
  \bibitem{Hume2007}
D.~B. Hume, T. Rosenband, and D.~J. Wineland, Phys. Rev. Lett. {\bf 99},
  120502  (2007).

\bibitem{Myerson2008}
A.~H. Myerson {\it et~al.}, Phys. Rev. Lett. {\bf 100},  200502  (2008).


\bibitem{Note2}
With the chosen thresholding parameters, only about 10\% of the data is
  discarded. A hypothetical bias introduced by this method can therefore only
  have a small influence on the measured detection fidelity.


\bibitem{Boehi2009}
P. B\"ohi {\it et~al.}, Nature Physics {\bf 5},  592  (2009).

\bibitem{Rosenfeld2009}
W. Rosenfeld {\it et~al.}, Advanced Science Letters {\bf 2},  469  (2009).

\end{thebibliography}

\end{document}